\documentclass{ws-p8-50x6-00}
\newcommand{\nat}{Nature,~ }
\newcommand{\apjl}{Ap. J. Lett.,~}
\newcommand{\apj}{Ap. J., ~}
\newcommand{\physrep}{Physics Reports,~}
\newcommand{\aaps}{Astron. \& Astroph. Supp.,~}

\newcommand{\mnras}{Mon. Not. RAS, ~}

\begin{document}

\title{Gamma-Ray Bursts - a Primer For Relativists}

\author{Tsvi Piran}

\address{Racah Institute for Physics, The Hebrew University, Jerusalem, 91904, Israel\\
E-mail: tsvi@phys.huji.ac.il}


\maketitle

\abstracts{Gamma-Ray Bursts (GRBs) - short bursts of 100-1MeV
photons arriving from random directions in the sky are probably
the most relativistic objects discovered so far. Still, somehow
they did not attract the attention of the relativistic community.
In this short review I discuss briefly GRB observations and show
that they lead us to the fireball model - GRBs involve
macroscopic relativistic motion with Lorentz factors of a few
hundred or more. I show that GRB sources involve, most likely, new
born black holes, and their progenitors are Supernovae or neutron
star mergers. I show that both GRB progenitors and the process of
GRB itself produce gravitational radiation and I consider the
possibility of detecting this emission. Finally I show that GRBs
could serve as cosmological indicators that could teach us about
the high redshift ($z \approx 5-15$) dark ages of the universe.}

\section{Introduction}

Gamma-Ray bursts - GRBs, short and intense bursts of
$\gamma$-rays arriving from random directions in the sky were
discovered accidentally more than thirty years ago. During the
last decade two detectors, BATSE on  CGRO and BeppoSAX have
revolutionized our understanding of GRBs. BATSE has demonstrated
\cite{BATSE92} that GRBs originate at cosmological distances in
the most energetic explosions in the Universe. BeppoSAX
discovered X-ray afterglow\cite{Costa97}. This enabled us to
pinpoint the positions of some bursts, locate optical
\cite{vanP97} and radio \cite{Frail97} afterglows, identify host
galaxies and measure redshifts to some bursts \cite{Metzger97}.

The high energy release and the rapid time scales involved
suggested very early that GRBs might be associated with
relativistic compact objects. The discoveries of BATSE and
BeppoSAX confirmed these expectations. These observations  have
established the Fireball model demonstrating that GRBs are the
most relativistic objects known so far:
\begin{itemize}
\item GRBs
involve macroscopic ultrarelativistic flows with Lorentz factors
$\Gamma \ge 100$.
\item
GRBs involve accretion onto a newborn compact object, most likely
a black hole. GRBs are the birth cries of these black hole.
\item GRBs could emit significant
amounts of gravitational radiation; both in the process of the
formation of the black hole and during the acceleration of the
matter to relativistic velocities.
\item GRB can serve as beacon lights that would help us to explore
the high redshift universe.
\end{itemize}

I review, here,   our understanding of GRBs, emphasizing, as
appropriate for this conference, their relativistic nature. I
begin in with a very brief exposition of the properties of GRBs.
I continue in section \ref{fireball} with a short exposition of
the Fireball model (see \cite{P99,P00} for details), focusing on
realization that GRBs must involve ultrarelativistic motion and
on the observational proofs of such motion. In \ref{engine} I
summarize the implication of the Fireball model to the ``inner
engines" and showing their association with black holes. I also
discuss the possible progenitors of GRBs. In section
\ref{gravrad} I discuss GRBs as sources of Gravitational
radiation.  I will not discuss here how GRBs could be used to
explore the early high z universe and I refer the reader to recent
reviews \cite{Lamb01,Lamb00} on this issue. Concluding remarks,
predictions and open questions are discussed in section
\ref{conclusions}.

\section{Gamma Ray Bursts Observations}

Gamma-Ray Bursts are short and intense bursts of soft gamma-rays,
arriving from random directions in the sky. Most observational
features of GRBs are nicely summarized in various reviews see e.g.
\cite{Fishman95,Paradijs00}. Some of the basic features of the
prompt burst are:
\begin{itemize}
\item The bursts originate from cosmological distances and arrive from random directions in the sky.
\item The overall observed fluences range from
$10^{-4}$ergs/cm$^2$ to $10^{-7}$ergs/cm$^2$ (the lower limit
depends of course on the characteristic of the detectors and not
on the bursts themselves). This corresponds to isotropic emission
of $10^{53}$ergs. However, we know today that most GRBs are
narrowly  beamed and the corresponding energies are around
$10^{51}$ergs \cite{Frail01,PanaitescuK01,Piran01}.
\item The spectrum is non thermal. The energy flux peaks at few
hundred keV and there is a long high energy tail extending in
cases up to GeV.
\item The duration of the bursts ranges from less
than 0.01sec to more than 100sec.  GRBs are divided to short
($T<2$sec) and long ($T> 2$sec) according to their duration.
\item The light curves show rapid variability, at times on scales
less than 10msec.

\item The present local rate of long observed GRBs is  $\approx ~2 {\rm Gpc}^{-3}
{\rm yr}^{-1}$ \cite{Schmidt99,Schmidt00}. The rate of GRBs seems
to follow the star formation rate, namely this rate was higher by
a factor of 10 at $z > 2$. The rate will of course be higher by a
factor $2 / \theta^2$,  if GRBs are beamed with an opening angle
$\theta$.
\item The rate of short bursts is uncertain. There are
indications that short bursts are weaker than long ones and hence
the observed short bursts are nearer to us that the long ones
\cite{Mao_narayan_piran,Katz}. The best estimate so far suggests
that all short bursts are at $z<0.5$.  So far afterglow was not
detected from any short burst and there is no redshift
measurement to any short burst hence there is no independent
confirmation of this estimate.  Given about 250 short bursts per
year and assuming that they all come from within $z<0.5$ we find
that the current rate of observed short bursts, $R_{iso-short}= 2
\times 10^{-8} {\rm Mpc}^{-3} {\rm yr}^{-1}$, is about ten times
larger than the rate of long GRBs.
\end{itemize}

Until 1997 there where no known counterparts to GRBs in other
wavelengths. On Feb 28 1997 the Italian-Dutch satellite  BeppoSAX
detected x-ray afterglow from GRB 970228. The exact position
given by BeppoSAX led to the discovery of optical afterglow
\cite{vanP97}. Radio afterglow was detected in GRB 970508
\cite{Frail97}. By now more than thirty x-ray afgterglows have
been observed. About half of these have optical and radio
afterglow and in many of those the host galaxy has been
discovered and in a dozen or so cases their redshift has been
measured. The observed redshifts range from 0.46 to a record of
4.5.

\section{Ultra-Relativistic Motion and the Fireball Model}
\label{fireball}

\subsection{The Compactness Problem and Ultra-Relativistic Motion}

The need for ultrarelativistic motion in GRBs arise from the
conflict between the large energy released, the short time scale
and the non thermal spectrum. With a fluence of $\sim
10^{-6}$ergs/cm$^{2}$ and a cosmological distance $\sim 10^{28}$cm
the energy of GRBs is $\sim 10^{51}$ergs. This energy is released
within a few seconds. Using the usual causality limit on the size
of an object, $R$, given a variability time scale, $\Delta T$, $ R
\le c \Delta T$ to estimate the density of photons one finds that
the optical depth for pair production $\gamma\gamma \rightarrow
e^+e^-$ would be $\sim 10^{15}$ (Piran, 1999). Such a source
cannot emit the observed nonthermal spectrum.

Already in 1975 Ruderman \cite{Ruderman75} pointed out that
relativistic effects could eliminate this problem. First if the
source is moving relativistically with a Lorentz factor $\Gamma$
towards us than the usual causality limit  is replaced by: $R \le
c \Delta T / \Gamma^2$. Additionally if the source is moving
towards us the observed photons have been blue shifted. At the
source they have lower energy which would be insufficient for
pair production. Together this leads to a decrease in the
estimated optical depth by a factor of $\Gamma^{2+2\alpha}$,
where $\alpha \sim 2 $ is the spectral index, namely the exponent
of the photon number distribution. Various estimates
\cite{FenimoreEH93,WoodsL95,P95,SaLi00}, of the Lorentz factor
$\Gamma$ based on the compactness problem lead to comparable
values, $\Gamma \ge  100$, (see \cite{SaLi00} for a critical
review). Today we have independent direct observational evidence
for such ultra-relativistic motion.

\subsection{The Fireball Model}

While we don't expect celestial objects to roam around at
ultra-relativistic velocities one can imagine spherical explosions
or jets in which matter is ejected from a central source
ultra-relativistically. This leads us to the Fireball model.

The Fireball model asserts that GRBs are produced when the
kinetic energy (or Poynting flux) of a relativistic flow is
dissipated by shocks. These shocks accelerate electrons and
generate strong magnetic fields. The relativistic electrons emit
the observed $\gamma$-rays via synchrotron or Synchrotron-self
Compton. There are two variants of this model: The External
Shocks model \cite{MR93a} assumes that the shocks are between the
relativistic flow and the surrounding circumstellar matter. The
Internal Shocks model \cite{NPP92,RM94} assumes that the flow is
irregular and the shocks take place between faster and slower
shells within the flow. The rapid time variability seen in most
GRBs cannot be produced by external shocks
\cite{SP97,Fenimore96}. This leaves internal shocks as the only
viable model for the production of the GRB. As internal shocks do
not dissipate all the kinetic energy, the remaining energy will be
dissipated later by interaction with the surrounding matter and
produce an afterglow. This leads us to the Internal-External
shocks model \cite{SP97}. According to this model Internal shocks
are responsible for the GRB while external shocks produce the
longer lasting afterglow (see Fig. 1). The shocks occur at
relatively large distances ($10^{13}-10^{14}$cm for internal
shocks and $10^{14}-10^{18}$cm for external shocks) from the
source that generates the relativistic flow. The observed
radiation from the GRB or from the afterglow reflects only the
conditions within these shocks. We have only indirect information
on the nature of the ``inner engines".

\begin{figure}[t]
\epsfxsize=28pc 
\epsfbox{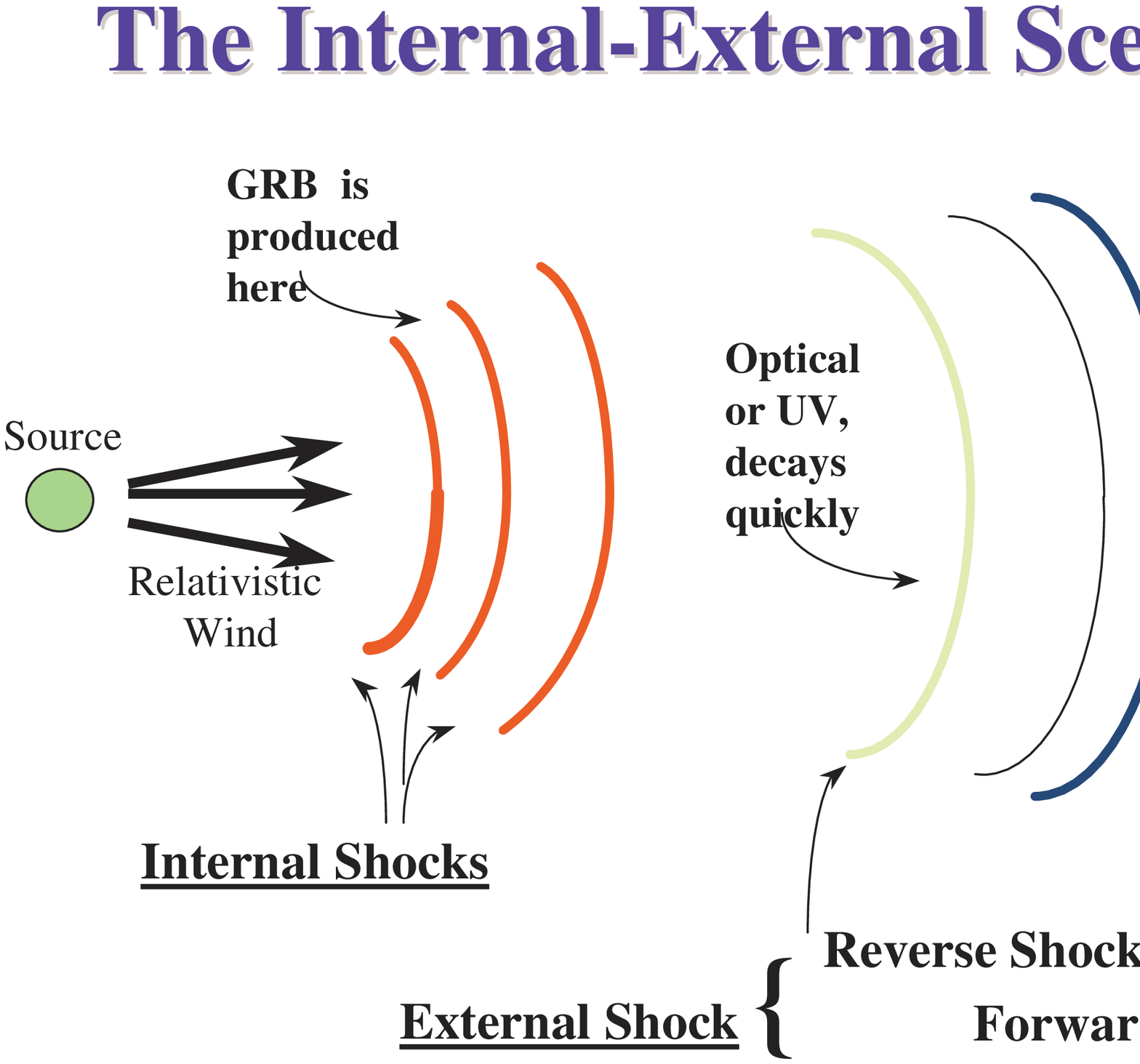} 
\caption{\bf The External Shocks Model.} \label{fig1}
\end{figure}

Fig. 1 depicts a schematic picture of the Internal-External
shocks model. An inner engine produces an irregular wind. The
wind varies on a scale $\delta t$ and its overall duration is
$T$. The variability scale $\delta t$ corresponds to the
variability scale observed in the GRB light curve \cite{KPS97}
thus, $\delta t \le 1$ sec. Internal shocks take place at $R
\approx \delta t \Gamma^2  \approx 3 \cdot 10^{14} {\rm cm}
~(\delta t/1~ sec) (\Gamma/100)^2$.  External  shocks become
significant at $\sim 10^{16}$cm (see \cite{SaP95,P99} for
details). Initially, there is also a short lived reverse shock
that propagates into the ejecta. This reverse shock is
responsible to the prompt optical emission observed in GRB990123
that we discuss later

The forward shock propagates into the surrounding matter
producing the afterglow. It turns out to a   blast wave that is
well described by the Blandford-McKee self similar solution. The
theory of the afterglow is well understood \cite{BM76}. Blandford
and McKee have worked out (already in the seventies!) the theory
of an adiabatic relativistic blast wave. This is the relativistic
analog of the well known Sedov-Taylor solution. Electrons are
accelerated to relativistic velocities by the shocks and their
interaction with the magnetic field leads to synchrotron
radiation. This provides an excellent model for the observed
emission \cite{SPN98}. Overall we have a simple theory
characterized by five parameters: the total energy, $E_0$, the
ambient density, $n_0$, the ratio of the electrons and magnetic
fields energy density to the total energy density, $\epsilon_e$,
$\epsilon_B$ and the exponent  of the electrons' energy
distribution function $p$. An additional sixth parameters, the
exponent of the circumstellar density distribution, $n$, arises
in cases when the external matter density is inhomogeneous ($\rho
\propto r^{-n}$). Most notable is $n=2$ corresponding to a
pre-GRB wind expected in some models \cite{Chevalier00}. This
rather simple theory predicts a robust relations between $\alpha$
and $\beta$ the exponents describing the flux as a function of
frequency, $F_\nu \propto t^{-\alpha} \nu^{-\beta}$. At the high
frequencies, above the cooling frequency, we have (for $n=0$),
$\alpha = (3p-2)/4$ and $\beta=p/2$.

\subsection{Observations of Relativistic Motion}

The radio afterglow observations of GRB 970508 provided the first
verification of relativistic motion. The radio light curve (in
4.86Ghz) varied strongly during the first month. These variations
died out later. Even before this transition Goodman
\cite{Goodman97} interpreted these variations as scintillations.
The observation of a transition after one month enabled Frail et
al., \cite{Frail97} to estimate the size of the afterglow at this
stage as $\sim 10^{17}$cm. It immediately follows that the
afterglow has expanded relativistically. Additionally, the source
is expected to be optically thick in radio \cite{KP97} leading to
a $\nu^2$ rising spectrum at these frequencies. The observed flux
from the source enables us (using the black body law) to estimate
the size of the source. As predicted the radio spectrum increases
like $\nu^2$. The size estimated with this method agrees
\cite{Frail97} with the one derived by the scintillations
estimate implying as well a relativistic motion.

The radio emission from  GRB970508 showed relativistic motion in
its afterglow. However, these observations took place one month
after the burst and at that time the motion was only ``mildly"
relativistic with a Lorentz factor of order a few. Are there
``direct" observations for the ultrarelativistic motion that
exists earlier?

In Jan 23 1999 ROTSE recorded six snapshots of optical emission
from GRB990123\cite{ROTSE} .  Three of those were taken while the
burst was still emitting $\gamma$-rays. The other three snapshots
spanned a couple of minutes after the burst. The second snapshot,
taken 70sec after the onset of the burst corresponds to a 9th
magnitude signal.  A comparison of these optical observations
with the $\gamma$-rays and X-rays light curves (see e.g.
\cite{P00})  shows that the optical emission does not correlate
with the $\gamma$-rays pulses. The optical photons and the
$\gamma$ rays are not emitted by the same photons
\cite{SP99b,ROTSE}.

How can one explain this flash? The collision between the ejecta
and the surrounding medium produces two shocks. The outer forward
shock propagates into the ISM. This shock develops later into the
self similar Blandford-McKee \cite{BM76} blast wave that drives
the afterglow. A second shock, the reverse shock,  propagates into
the flow. This reverse shock is short lived. It dies out when it
runs out of matter as it reaches the inner edge of the flow.
While it is active, it is a powerful source of energy. Comparable
amounts of energy are dissipated by the forward and by the
reverse shocks \cite{SaP95}.  Sari and Piran \cite{SP99c}
predicted (just a few month before GRB990123) that this reverse
shock will produce an intense (brighter than 11th magnitude)
prompt optical flash.

The observations of an optical flash demonstrated a relativistic
motion with $\gamma \sim 200$ during the burst. There are three
independent estimates of the Lorentz factor at the time that the
ejecta hits first the ISM \cite{SP99b}. First the time delay
between the GRB and the optical flash suggests $\Gamma \sim 200$.
The ratio between the emission of the forwards shock (x-rays) and
the reverse shock (optical) gives another estimate of $\Gamma
\sim 70$. Finally the fact that the maximal synchroton frequency
of the reverse shock was below the optical band led to $\Gamma
\sim 200$. The agreement between these three crude and
independent estimates is reassuring.

\subsection{Jets, Beaming and Energetics}

With redshift measurements it became possible to obtain exact
estimates of the total energies involved. While the first burst
GRB970508 required a modest value of $\sim 10^{51}$ergs, the
energies required by other bursts were alarming, $3 \times
10^{53}$ergs for GRB981226 and $4 \times 10^{54}$ergs for
GRB990123, and unreasonable for any simple compact object model.
These values suggested that the  assumed isotropic emission was
wrong and GRBs are beamed. Significant beaming would of course
reduce, correspondingly the energy budget.

Beaming was suggested even earlier as it arose naturally in some
specific models. For example the binary neutron star merger has a
natural funnel along its rotation axis and one could expect that
any flow would be emitted preferably along this axis. The
Collapsar model, in which the GRB is produced during the collapse
of a massive star, also requires beaming. Only a concentrated
beamed energy could drill a whole through the stellar envelope
that exists in this model.

Consider a relativistic flow with an opening angle $\theta$. As
long as $\theta > \Gamma^{-1}$ the forwards moving matter doesn't
``notice" the angular structure  and the hydrodynamics is
``locally" spherical \cite{P94}. The radiation from each point is
beamed into a cone with an opening angle $\Gamma^{-1}$.  It is
impossible to distinguish at this stage a jet from a spherical
expanding shell. When $\theta \sim \Gamma^{-1}$ the radiation
starts to be beamed sideways. At the same time the hydrodynamic
behaviour changes and the material starts expanding sideways.
Both effects lead to a faster decrease in the observed flux,
changing $\alpha$, the exponent of the decay rate of the flux to:
$\alpha=p/2$. Thus we expect a break in the light curve and a new
relation between $\alpha$ and $\beta$ after the break
\cite{Rhoads99,SPH99,PM99}. The break is expected to take place
at $t_{jet} \approx 6.2 (E_{52}/n_0)^{1/3} (\theta/0.1)^{8/3}$hr
\cite{SPH99}. The magnitude of the break and the time of the
transition are different if the jet is expanding into a wind with
$r^{-2}$ density profile \cite{PanaitescuK00}. Recently numerical
simulation \cite{Granot01} have shown that the break appears in a
more realistic calculations, even though the numerical results
suggest that the analytical model developed so far are probably
too simple.

GRB980519 had unusual values for $\alpha=2.05$ and $\beta=1.15$.
These values do not fit the "standard" spherical afterglow
model\footnote{A possible alternative fit is to a wind (n=2)
model but with a unsual high value of $p=3.5$}. However, these
values are in excellent agreement with a sideway expanding jet
\cite{SPH99}. The simplest interpretation of this data is that we
observe a jet during its sideway expansion phase (with $p=2.5$).
The jet break transition from the spherical like phase to this
phase took place shortly after the GRB and it was not caught in
time. The light curves of GRB990123 shows, however, a break at
$t\approx 2$days \cite{Kulkarni99}. This break is interpreted as
a jet break, corresponding to an opening angle $\theta \sim 5^o$.
Another clear break was seen in GRB990510
\cite{Harrison99,Stanek99}.

The brightest bursts, GRB990123 and GRB980519 gave the first
indications for jet like behaviour \cite{SPH99}. This suggested
that their apparent high energy was due to the narrow beaming
angles. A compilation of more bursts with jet breaks suggests
that all bursts have a comparable energy $\sim 10^{51}$ergs and
the variation in the observed energy is mostly due to the
variation in the opening angles $\theta$
\cite{Frail01,PanaitescuK01,Piran01}.

\subsection{Variability, Internal Shocks and a ``NO GO Theorem"}

According to the internal shocks model the observed light curve
corresponds to the temporal activity of the ``inner engine"
\cite{KPS97}. Further indication for this understanding arise
from the observations that the distribution of pulse width and
pulse separations are similar and that pulse widths are
correlated with the intervals preceding of following them
\cite{NakarP01a}.

These results imply that there must be two  different time scales
operating within the ``inner engine". The short time scale is the
variability time scale. The duration of the observed burst
corresponds to the time that the ``inner engine" is active. This
time scale  is up to 5 orders of magnitude longer then the short,
variability scale. This leads us to a NO GO Theorem. One cannot
produce a variable GRB by a single explosion in which all the
energy is released at once. This NO GO theorem rules out dozens
of GRB models (such as evaporating black holes, vaccum
instability, transition from a neutron star to a strange star
etc...) which involve sudden energy release.

\section{Black Holes,  the Inner Engines and GRB Progenitors}
\label{engine}

The Fireball model tells us how GRBs operate. However, it does
not answer the most interesting astrophysical questions: what
produces them and how? I turn to this issues now.

GRBs must involve compact objects. There is no other way to
extract so much energy, $\sim 10^{51}$ergs, so quickly. We have
seen earlier that the temporal behaviour puts some stronger
limits on the source. The short time scale, which is as short as
a few ms, also suggest that we are dealing with a  compact
object. The long duration of the bursts shows that the source
must be active much longer than its dynamical time scale. This
suggests that GRBs arise due to accretion and this time scale is
the duration of the accretion process. The energy involved
requires a massive ($\sim 0.1-0.2 m_\odot$) disk. Such a massive
disk can form only as debris during the formation of the compact
object itself. With such a massive disk the most likely compact
object is therefore a newborn black hole. A black hole is also
the natural consequence of the two most common progenitors:
Collapsars and neutron star mergers.

Accretion is needed to produce the two different time scales, and
in particular the prolonged activity. A massive ($\sim 0.1
m_\odot$) disk is required because of the energetics. We expect
that such a massive disk can form only simultaneously with the
formation of the compact object. This leads to the conclusions
that GRBs accompany the formation of black holes. This model is
supported by the observations of relativistic (but not as
relativistic as in GRBs) jets in AGNs, which are powered by
accretion onto black holes. This system is capable of generating
collimated relativistic flows even though we don't understand how.

An important alternative to accretion is Usov's model
\cite{Usov92} in which the relativistic flow is mostly Poynting
flux and it  is driven  by the magnetic and rotational energies
of a newborn rapidly rotating neutron star. However  this model
seems to fall short by an order of magnitude of the energy
required. Additionally, there is no indication of the slowing down
pattern (that would be expected in such a case) in the light
curves of GRBs.

\subsection{GRB progenitors}

Several scenarios could lead to a black hole - massive accretion
disk system. This could include mergers (NS-NS binaries
\cite{Eichler_LPS89,NPP92}, NS-BH  binaries \cite{Pac91}  WD-BH
binaries \cite{Fryer_WHD99}, BH-He-star binaries
\cite{Fryer_Woosley98}) and models based on ``failed supernovae''
or ``Collapsars'' \cite{Woosley93,Pac98,MacFadyen_W99}. Narayan
et al. \cite{NarayanPK01} have recently shown that accretion
theory suggests that from all the above scenarios only Collapsars
could produce long bursts and only NS-NS (or NS-BH) mergers could
produce short bursts.


Additional  indications on the astrophysical nature of the
sources arise from afterglow observations. One has to use these
clues with care. Not all GRBs have afterglow (for example,  so far
afterglow was not detected from any short burst) and  it is not
clear whether these clues are relevant to the whole GRB
populations. These clues seem to suggest a GRB-SN connection:
   \\ \indent{\bf $\bullet$ SN association:}
   Possible association of GRB980425 with SN98bw \cite{GalamaSN98}
  and  possible SN signatures in the afterglows of
  GRB970228 \cite{Reichart99} and GRB980326 \cite{Bloom99}.
   \\ \indent{\bf $\bullet$ Iron lines:} have been observed
  in some x-ray afterglows \cite{Piro00}. Any model explaining them requires a
  significant amounts of iron at rest near those GRBs.
     \\ \indent {\bf $\bullet$ Association with Star formation:}
     GRBs seem to follow the star formation rate.
  GRB are located within  star forming
  regions in star forming Galaxies \cite{Pac98}.
     \\ \indent {\bf $\bullet$ GRB distribution:}
     GRBs are distributed within galaxies. There is no evidence for
  GRBs kicked out of their host galaxies  \cite{Bloom01}
  as would be expected for NS-NS mergers  \cite{NPP92}.

All these clues  point out towards a SN/GRB association and
towards the Collapsar model. However, the situation is not clear
cut. The association of GRB980425 with SN98bw is uncertain. There
are alternative explanations to the bumps in the afterglows of
GRB970228 and GRB980326 \cite{Esin00}. Iron is produces in
Supernovae. But there is no simple explanation what is iron at
rest doing around the GRB (A model that explains the formation of
iron lines requires that the supernova took place several month
before the burst \cite{Vietri01}. This would be incompatible with
the reported SN bumps which coincide with the GRB). One should
bear in mind that association with star formation and the
distribution of GRBs within galaxies indicates that GRB stellar
progenitors are short lived. This is compatible with massive
stars. However, one cannot rule out a short lived binary NS
population \cite{TutukovY94} which would mimic this behaviour.

We stress that there are some indication that seem incompatible
with the SN association:
     \\ \indent {\bf $\bullet$ No Windy Afterglow:} No evidence for a wind (n=2)
  in any of the afterglow light curves. Such winds are expected from massive progenitors.
  Furthermore,
  most fits for the afterglow parameters show low
  ambient density \cite{PanaitescuK01}.
     \\ \indent{\bf $\bullet$ No Jets:} Some GRBs don't show evidence for a jet or have very
  wide opening angles \cite{PanaitescuK01}, this would be incompatible
  with the Collapsar model.

\section{Gravitational Radiation from GRBs}
\label{gravrad}
 The appearance of relativistic nonspherical
(jets) motion and the association with black holes suggests that
GRBs are be potential sources of gravitational radiation. There
are two phases in which gravitataional radiation can arise in
GRBs. First from the process that lead to the formation of the
black hole and second from the fireball process itself.

I consider first gravitational radiation that arises before the
GRB, as part of the formation of the ``inner engine". Here I
consider the two main progenitors candidates: Collapsar and
neutron star mergers.

\subsection{Mergers}

I consider here both binary neutron star mergers and black
hole-neutron star mergers under the single category of mergers.
These sources are the ``canonical" sources of gravitational
radiation emission.  Both LIGO and VIRGO  aim in detecting these
sources. Specifically the goal of these detectors is to detect
the characteristic``chirping" signals arising from the
in-spiraling phase of these events. The possibility of detection
of such signals has been extensively discussed (see e.g.
\cite{LIGO-merger}) and we won't repeat this here. Such events
could be detected up to a distance of $\sim 20$Mpc with LIGO I
and up to $\sim 300-600$Mpc with LIGO II.

The detection of the chirping merger signal is based on fitting
the gravitational radiation signal to pre-calculated templets.
Kochaneck and Piran \cite{Kochaneck_Piran93} suggested that the
detection of a merger gravitational radiation signal would
requite a lower S/N ratio if this signal coincides with a GRB.
This could increase somewhat the effective sensitivity of LIGO
and VIRGO to such events.

It is expected that mergers (either binary neutron star or a black
hole-neutron star mergers) produce the short GRBs (see
\cite{NarayanPK01}).  Considering the isotropic rate of short GRBs
estimated earlier we find that there should be one short burst per
year within $\sim 450$Mpc.  This is just at the sensitivity level
of LIGO II.  As already mentioned it is not clear if short GRBs
are beamed.  If they are beamed, with the same beaming factor as
long GRBs we should expect several hundred mergers events per a
single observed burst.  This would put one merger event per year
at $\sim 80\theta_{0.1}^2$Mpc.

The corresponding distances to long GRBs are much longer.  The
nearest (long) GRB detected within a year would be a t 1Gpc. This
is far beyond the sensitivity of even LIGO II.  Beaming puts the
nearest (long) event much nearer, at $135 \theta_{0.1}^2$Mpc,
well within the sensitivity of LIGO II. However, this burst would
be, most likely, directed away from us and won't be observed as a
GRB. Still a GRB that is beamed away from us is expected to
produce an ``orphan" afterglow and the gravitational radiation
signal could trigger a search for this afterglow.

\subsection{Collapsars}

The Collapsar model \cite{Woosley93,MacFadyen_W99} is based on
the collapse of the core of a massive star to a black hole
surrounded by a thick massive accretion disk. The accretion of
this disk onto the black hole, is accompanied by the acceleration
of ultra relativistic jets along the rotation axis and powers the
GRB. The jets first have to punch a hole in the stellar envelope.
The GRB forms only after the jets have emerged from the envelope.
Due to the relatively long time that it takes for the jets to
punch a hole in the envelope it is expected that Collapsars can
produce only long bursts.

As far as gravitational radiation is concerned this system is very
similar to a regular supernova.  Rotating gravitational collapse
has been analyzed by Stark and Piran \cite{Stark_Piran85}. They
find that the gravitational radiation emission emitted in a
rotating collapse to a black hole is dominated by the black
hole's lowest normal modes, with a typical frequency of $\sim
20c^3/GM$. The total energy emitted is:
\begin{equation}
{\Delta E_{GW}} = \epsilon M c^2 = \rm{min}(1.4 \cdot 10^{-3}
a^4, \epsilon_{max})   M c^2 \ ,
\end{equation}
where $a$ is the dimensionless specific angular momentum and
$\epsilon_{max}$ is a maximal efficiency which is of the order
${\rm a~~ few} \times 10^{-4}$.  The expected amplitude of the
gravitational radiation signal, $h$, would be of the order of
$\sqrt{\epsilon} GM/c^2 d$ where $d$ is the distance to the
source.  Even LIGO II won't be sensitive enough to detect such a
signal from a distance of 1Gpc or even from 100 Mpc. Furthermore,
this signal would be rather similar to a supernova gravitational
radiation signal.  As regular supernovae are much more frequent
it is likely that a supernova gravitational radiation signal
would be discovered long before a Collapsar gravitational
radiation signal.

\subsection{Gravitational Radiation Emission from the GRB Itself}

I turn now to examine the gravitational radiation that would
arise from the GRB process itself. According to the fireball
model the ``inner engine" accelerates a mass of $M = E/\Gamma c^2$
to a Lorentz factor $\Gamma$. The most efficient generation of
gravitational radiation could take place here during the
acceleration phase, in which the mass is accelerated to a Lorentz
factor $\Gamma$. To estimate this emission we follow Winberg's
\cite{Weinberg73} analysis of gravitational radiation emitted
from a relativistic collision between two particles.

I consider the following simple toy model.  Two particles at rest
with a mass $M$ are accelerated instantly at $t=0$ to a Lorentz
factor $\Gamma$ and energy $E$. Conservation of energy requires
that some (actually most) of the rest mass was converted to
kinetic energy during the acceleration and the rest mass of the
accelerated particle is $m = E/\Gamma = M/\Gamma$. Using the
formalism developed by Weinberg \cite{Weinberg73} to estimate the
gravitational radiation generated in particle collisions, we
calculate the gravitational radiation emitted by this system.
Prior to the acceleration the two particles has momenta $m_0 (1,
0,0,0)$. After the acceleration the particles' momenta is $m
\Gamma (1, \pm \vec \beta)$.  The energy  emitted per unit
frequency per unit solid angle in the direction at an angle
$\alpha$ relative to $\vec \beta$ is:
\begin{equation}
{d E \over d \Omega d \omega} = {G M^2 \beta^2 \over c \pi^2}
\big[   {\Gamma^2 (\beta^2 - \cos^2\alpha) \over  (1 - \beta^2
\cos^2\alpha)^2} + { \cos^2\alpha \over \Gamma^2 (1 - \beta^2
\cos^2\alpha)^2} \big]\ . \label{flux}
\end{equation}
The result is independent of the frequency, implying that the
integral over all frequency  diverges. This nonphysical divergence
arises from the nonphysical assumption that the acceleration is
instantaneous. In reality this acceleration takes place over a
time $\delta t$, which is of order 0.01sec. This would produce a
cutoff $\omega_{max} \sim 2 \pi / \delta t$ above which Eq.
\ref{flux} is not valid. The angular distribution found in Eq.
\ref{flux} is dissppointing. The EM emission from the
ultrarelativistic source is beamed forwrds into a small angle
$1/\Gamma$, enhancing the emission in the forwards direction by a
large factor ($\Gamma^2$). We find here that the gravitational
radiation from this relativistic ejecta is spread rather
uniformly  in almost all $4\pi$ directions. Instead of beaming we
have ``anti-beaming" with no radiation at all emitted within the
forward angle $\Gamma^{-1}$ along the direction of the
relativistic motion.

\begin{figure}[t]
\epsfxsize=28pc 
\epsfbox{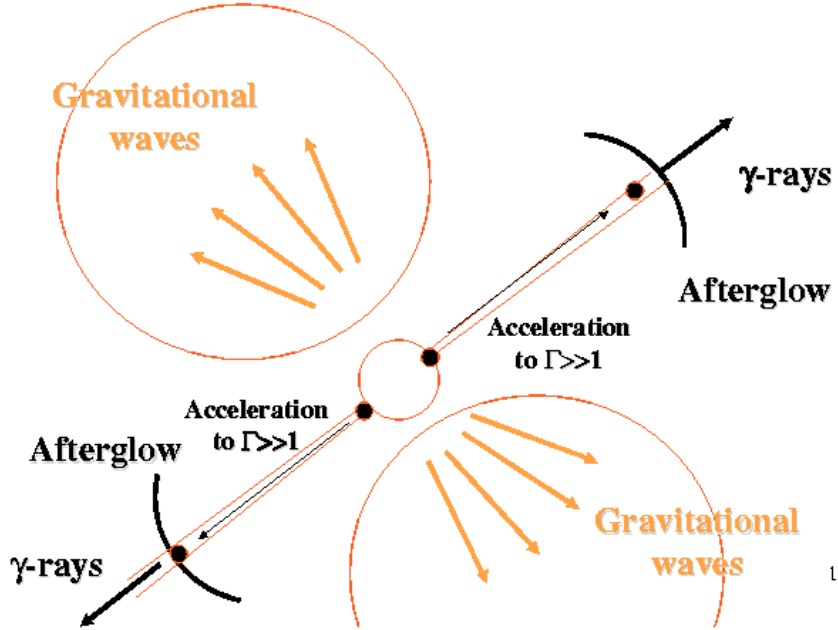} 
\caption{\bf Gravitational waves are mostly orthogonal to the
direction of the GRB and the afterglow.} \label{fig2}
\end{figure}

Integration of the energy flux over different directions yields:
\begin{equation}
{d E \over d \omega} = {G M^2 \over c \pi^2} [ 2 \Gamma^2 + 1+
{( 1 - 4 \Gamma^2) \over \Gamma^2 \beta} ArcTang(\beta)] \
.\label{energy_flux}
\end{equation}
As expected the total energy emitted is proportional to $m^2
\Gamma^2$. Further integration over frequencies up to the cutoff
$2 \pi / \delta t$ yields:
\begin{equation}
E  \approx { 2 G M^2 \Gamma^2 \over c \pi \delta t } \ .
\end{equation}

In reality the situation is much more complicated than the one
presented here. First, the angular width of the emitted blobs is
larger than $\Gamma^{-1}$. The superposition of emission from
different directions washes out the no emission effect in the
forward direction. Additionally according to the internal shocks
model the acceleration of different blobs go on independently.
Emission from different blobs should be combined to get the actual
emission. Both effects {\it reduce} the effective emission of
gravitational radiation and makes the above estimate an upper
limit to the emission that is actually emitted.

The gravitational signal is spread in all directions (apart from
a narrow beam along the direction of the relativistic motion the
GRB). It ranges in frequency from $0$ to $f_{max} \approx 100$Hz.
The amplitude of the gravitational radiation signal at the
maximal frequency, $f_{max} \approx 100$Hz, would be: $ h \approx
(GM\Gamma^2 /c^2 d) $.  For typical values of $E=M\Gamma =
10^{51}$ ergs, $\delta t = 0.01$ sec and a distance of $500 Mpc$,
we find $ h \approx .5 \times 10^{-25}$. This is far below the
sensitivity of planned gravitational radiation detectors. Even if
we consider a burst which is ten times nearer this "direct"
gravitational radiation signal would still be undetectable.

\section{Conclusions, Predictions and Open Questions}
\label{conclusions}

There is  an ample observational  support For the Fireball model.
Still there are many open questions. The most interesting ones,
from the point of view of this conference is how does the black
hole based, inner engine, operate. What is the energy source. How
does it convert the energy to ultra-relativistic flow and how does
it collimate the flow to narrow jets. These interesting issues
might be related to the Blandford-Znaek mechanism of extracting
energy from a rotating black hole via magnetic processes
\cite{BlandfordZnajek,Lee01,vanPutten00,LeeWijersBrown00}. It is
a unique challenge to relativists to explore the electrodynamics
of black hole and determine under what conditions this process can
operate effectively.

We know how GRBs are produced. We are less certain what produces
them. We can trace backwards the evolution at the source from the
observations of the emitting regions to an accretion disk -
black  hole system. The traces from this point backwards are less
clear. Theoretical considerations \cite{NarayanPK01} suggest that
only Collapsars can produce the disk-black hole systems needed for
long bursts while only NS-NS (or possibly NS-BH) mergers can
produce the systems needed for short bursts. These conclusions
are supported by afterglow  observations that suggest SN/GRB
association for the long burst population. However, the picture
is far from clear yet.

\section*{Acknowledgments}
This research was supported by a US-Israel BSF grant.


\end{document}